\documentclass[aps,prl,reprint,groupedaddress]{revtex4-2}
\usepackage{graphicx}
\usepackage{natbib}

\usepackage[table]{xcolor}
\usepackage{colortbl}
\usepackage{booktabs}
\usepackage{caption}
\usepackage{array}
\usepackage{tikz}
\usepackage{multirow}

\definecolor{c00}{rgb}{1.00, 0.00, 0.00} % red
\definecolor{c14}{rgb}{0.90, 0.20, 0.20}
\definecolor{c16}{rgb}{0.85, 0.25, 0.25}
\definecolor{c24}{rgb}{0.75, 0.35, 0.35}
\definecolor{c28}{rgb}{0.70, 0.40, 0.40}
\definecolor{c34}{rgb}{0.60, 0.50, 0.50}
\definecolor{c36}{rgb}{0.55, 0.50, 0.55}
\definecolor{c45}{rgb}{0.45, 0.55, 0.60}
\definecolor{c46}{rgb}{0.44, 0.55, 0.62}
\definecolor{c65}{rgb}{0.30, 0.60, 0.75}
\definecolor{c66}{rgb}{0.28, 0.60, 0.78}
\definecolor{c69}{rgb}{0.25, 0.65, 0.80}
\definecolor{c74}{rgb}{0.20, 0.70, 0.85}
\definecolor{c75}{rgb}{0.18, 0.72, 0.88}
\definecolor{c82}{rgb}{0.10, 0.80, 0.95}

\newcommand{\cellrb}[2]{%
  \ifdim #1pt < 0.5pt
    \cellcolor{#2}#1%
  \else
    \cellcolor{#2}#1%
  \fi
}

\begin{document}

%\preprint{}

\title{Coupled Interfacial Phenomena Suppress Propulsion in Catalytic Janus Colloids}


\author{Muhammad Haroon}
%\thanks{}
\author{Christopher Wirth}
%\email[]{clw22@case.edu}
\email{clw22@case.edu}
\affiliation{Department of Chemical and Biomolecular Engineering, Case Western Reserve University, Cleveland, OH 44106}


\date{\today}

\begin{abstract}
 Platinum-coated polystyrene Janus particles exhibit a combination of stochastic and deterministic motion in hydrogen peroxide solutions, making them promising candidates for applications in micro-scale cargo transport, drug delivery, and environmental remediation. The dynamics of Janus particles very near a boundary are dictated by conservative and non-conservative interactions that depend on particle, substrate, and solution properties. This study investigated the influence of orientational quenching by measuring the effect of changes in cap thickness and hydrogen peroxide concentration on particle velocity and maximum displacement. Janus particles with cap thicknesses of  3 nm, 7 nm, 10 nm, 20 nm, and 35 nm were analyzed in 1 wt./vol.\% and 3 wt./vol.\% hydrogen peroxide solutions near the bottom and top boundaries of the fluid cell. Results indicated that particles with lower cap thicknesses exhibit higher velocities, with faster particles in 3 wt./vol.\% peroxide as compared to 1 wt./vol.\% peroxide. Furthermore, results suggest a combined influence of activity and gravitational effects influenced whether particles moved along the top boundary i.e. ceiling or bottom boundary i.e. flooring. Heavier cap particles in lower peroxide concentration solution show less ceiling than lighter cap particles in higher peroxide concentration. We also find a global reduction in velocity for when a single surface of the two is plasma cleaned surface. These findings highlight the important interplay between cap weight, hydrodynamic interactions, and propulsion force in determining the dynamics of Janus particles.
\end{abstract}

%\keywords{}

\maketitle

\textit{Introduction} - Active matter is classified as a system that extracts energy from its environment to exert force and exhibit motion\cite{ramaswamy_active_2017}. Examples of active matter include microscopic living organisms, robots, and motors\cite{RevModPhys.85.1143}, all of which are typically at low Reynolds numbers $(Re \approx  10^{-6})$ where viscous forces dominate inertia. Often complementing the active force is that of stochastic "Brownian" fluctuations in both position and orientation that are relevant considering deterministic active motion. Various externally applied or locally generated fields, including electric\cite{ma_inducing_2015}, magnetic \cite{Tottori_Magnetic_2012}, optical\cite{buttinoni_active_2012}, acoustic  \cite{shields_evolution_2017}, thermal\cite{peng_opto-thermoelectric_2020}, or chemical gradients \cite{simmchen_topographical_2016}, are used to actuate motion.  In addition to serving as excellent experimental models for condensed matter physics, active particles are thought to have potential in diverse applications, including environmental remediation, micro-scale cargo transport\cite{sundararajan_catalytic_2008,baraban_transport_2012,balasubramanian_micromachine-enabled_2011}, microsurgery \cite{xi_rolled-up_2013}, bio-sensing \cite{bunea_sensing_2015}, water purification \cite{soler_self-propelled_2013}, oil recovery \cite{ tohidi_application_2022}, and drug delivery \cite{xu_sperm-hybrid_2018}. 
Active particles also serve as model systems for studying biological propulsion\cite{nsamela_colloidal_2023}. In both fundamental studies and applied uses of active particles, interactions involving the particle and nearby boundaries play a critical role. For instance, microorganisms often collect near interfaces (e.g. ocean vegetation or solid surfaces) making it essential to understand how boundaries influence their motion.\cite{ramia_role_1993}.

\begin{figure}
\includegraphics[width=\linewidth]{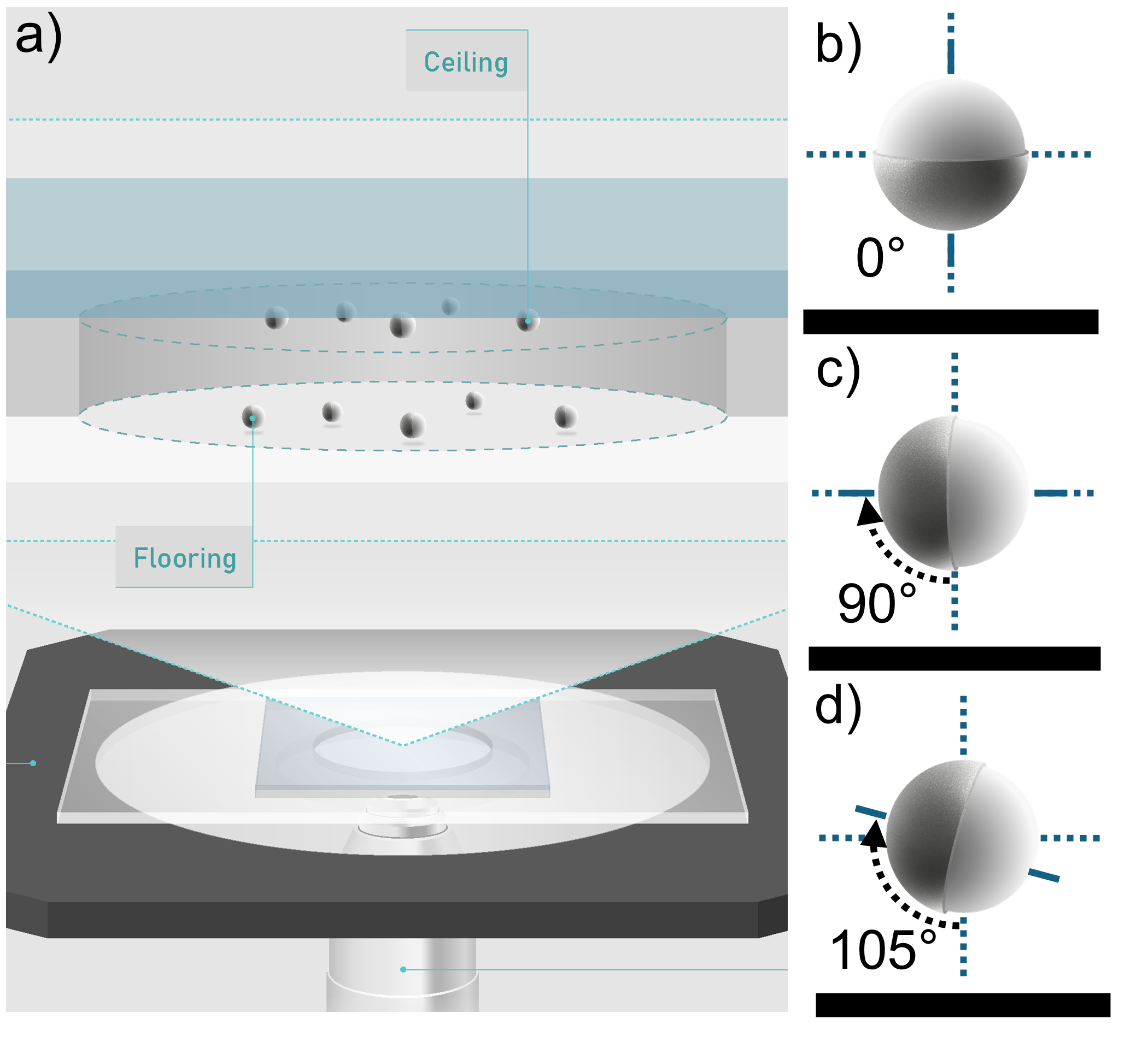}
\caption{\label{fig:setup} a) Experimental setup showing fluid cell on a microscope stage with ceiling and flooring states of particles moving in the fluid cell zoomed and labelled. b) Orientation of particle cap with respect to bottom boundary at 0 degrees. c) 90 degrees d) 105 degrees tilt.}
\end{figure}
Catalytically active Janus particles are one well studied example of an active matter system. These particles consist of a micrometer scale native sphere (often made of polystyrene or silica) having one hemisphere coated with a thin (approx. 10 nm) platinum film\cite{howse_self-motile_2007}. These particles self-propel in hydrogen peroxide solutions, where platinum catalyzes peroxide decomposition into water and oxygen, generating a propulsion force directed towards the polystyrene side. Increasing hydrogen peroxide concentration (the "fuel") increases propulsion velocity, whereas increasing particle size results in a reduction in particle velocity\cite{howse_self-motile_2007,ebbens_size_2012}. Typically, 1–10\% peroxide solutions with particles of 1–5 $\mu$m diameter exhibit velocities between 0.5–10 $\mu$m/s\cite{das_floor-_2020}.  The most prominent mechanisms cited as causing propulsion are those of ionic self-diffusophoresis and self-electrophoresis\cite{brown_ionic_2014,wang_open_2023}, although the dominant mechanism by which motion arises and at what conditions remains unclear. 

An emerging theme among studies over the past decade has been the importance of interfacial properties of either the particle itself or nearby boundary.  Proximity and orientation with respect to a boundary, as well as interfacial chemistry, have been shown to significantly impact propulsion speed. For example, Janus particles near boundaries tend to move in a sliding state due to hydrodynamic interactions, gravitational torque on cap, and sedimentation  \cite{das_floor-_2020,das_boundaries_2015,simmchen_topographical_2016}. Moreover, the angle at which an active particle approaches a boundary can effect whether it moves along the boundary, is reflected back, or becomes stationary\cite{mozaffari_self-diffusiophoretic_2016,uspal_self-propulsion_2014}. Particles can slide along both bottom and top boundaries of a fluid cell, referred to as flooring and ceiling states respectively, as shown in Figure~\ref{fig:setup} a \cite{das_floor-_2020}. Interesting recent work in this area has shown that altering the tilt angle of the platinum cap with respect to the boundary using an external electric field increases the particle velocity\cite{xiao_synergistic_2020}. Boundary chemistry can also influence propulsion. For example, particles move faster near hydrophobic surfaces than hydrophilic ones, due to osmotic coupling with the boundary \cite{Slip_length_Ketzetzi_2020}.  Note that plasma treatment is regularly used as a standard cleaning protocol for cleaning the substrate for Janus particle experiments, known to make the surface hydrophilic and negatively charged \cite{das_boundaries_2015}, which suggests that surface preparation may play a significant role in many of the existing studies. 

Complementing the boundary mediated phenomena described above is the geometry (coverage and thickness) and composition (material) of the cap itself. Modeling of non-active systems suggest that increasing cap thickness enhances the likelihood of a quenched state (cap pointing downward), where gravitational torque from the denser platinum (22.50 g/cm³) compared to polystyrene (1.05 g/cm³) suppresses rotational diffusion\cite{rashidi_influence_2020, rashidi_motion_2017, rajupet_derjaguin-landau-verwey-overbeek_2021, bottom-heavy-dynamics-Jalilvand-2025} . There has been considerably fewer studies that consider the effect of cap thickness in active systems \cite{mozaffari_self-diffusiophoretic_2016}. Yet, the modeling of non-active systems referenced above showed that even caps of moderate thickness will significantly alter the orientational dynamics of a Janus particle. One study showed that particle velocities decrease when the cap thickness is $<$5 nm due to lower coverage of the catalytic material on the particle surface, thus lowering the overall reaction rate on the platinum cap\cite{meng_effect_2022}. Several experimental studies have used different cap thicknesses between 5 nm and 20 nm and it is reported that platinum thickness is typically non-uniform, forming a crescent-like shape  \cite{das_floor-_2020,rashidi_local_2018,issa_engineered_2023,kalil_influence_2021, pawar2009multifunctional}. Theoretical predictions agree with previous experimental observations of caps orienting towards the bottom boundary in the absence of hydrogen peroxide in solution \cite{simmchen_topographical_2016}.  Upon addition of peroxide, particles slide in parallel along the bottom and top surfaces, slightly tilted away from the boundary \cite{wang_tilt_2025}. The average tilt of active particles is found to be approximately 105 $^{{\circ}}$ as illustrated in Figure~\ref{fig:setup} d. Further, it is shown that bottom-heaviness of a platinum cap can cause active particles to propel against gravity\cite{campbell_gravitaxis_2013}.  While Wang et al. showed velocity increases with increasing platinum cap thickness from 5 nm to 20 nm, their study did not specify whether measurements were conducted near a boundary. Further, the effect of cap thickness was outside of the scope of that initial study, whereas it instead explored differences arising from the surface contour of platinum in the cap (e.g. smooth vs. a nanoparticle platinum film) \cite{lyu_active_2021}. A more recent study has shown the velocities of a Janus particle near an indium-tin oxide (ITO) coated glass slide and in 5 v/v. \% peroxide will increase to as much as 5 $\mu$m/s  \cite{wang_tilt_2025}. One of the clever strategies in their work was to apply an intermittent A/C electric field, which was used to help control particle orientation. However, it was unclear what impact application of en electric field would have on equilibrium or transient charge on the particle or ITO surfaces.

\begin{figure}
\includegraphics[width=\linewidth]{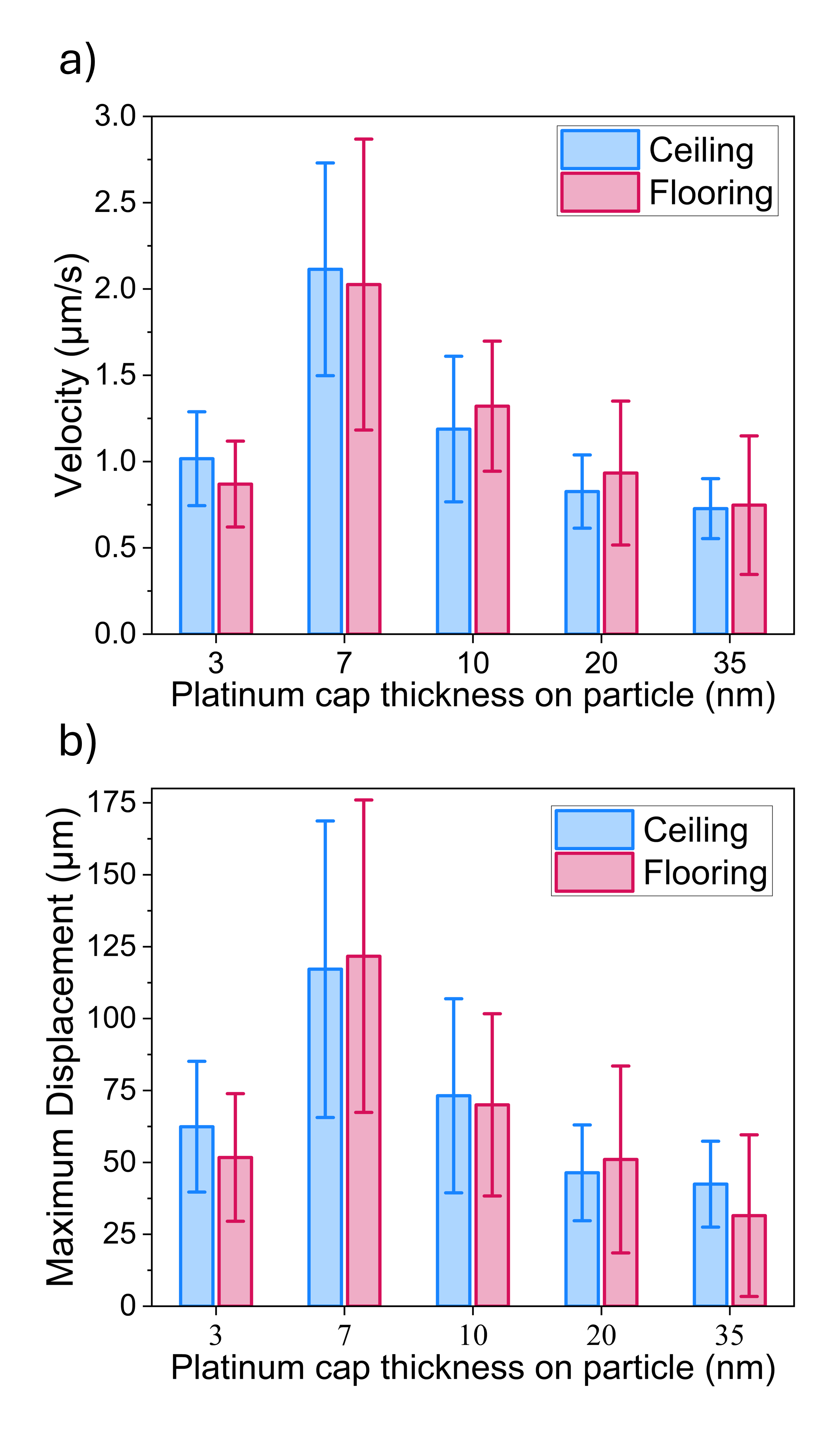}%
\caption{\label{FIG. 2.} a) Velocities and b) maximum displacements of particles of different cap thickness in  3 wt./vol \% hydrogen peroxide in water solution. Ceiling is shown in blue and flooring is show in red. Bar represents mean and error bars show +/- 1 standard deviation.}
\end{figure}

\begin{figure*}
\includegraphics[width=\linewidth]{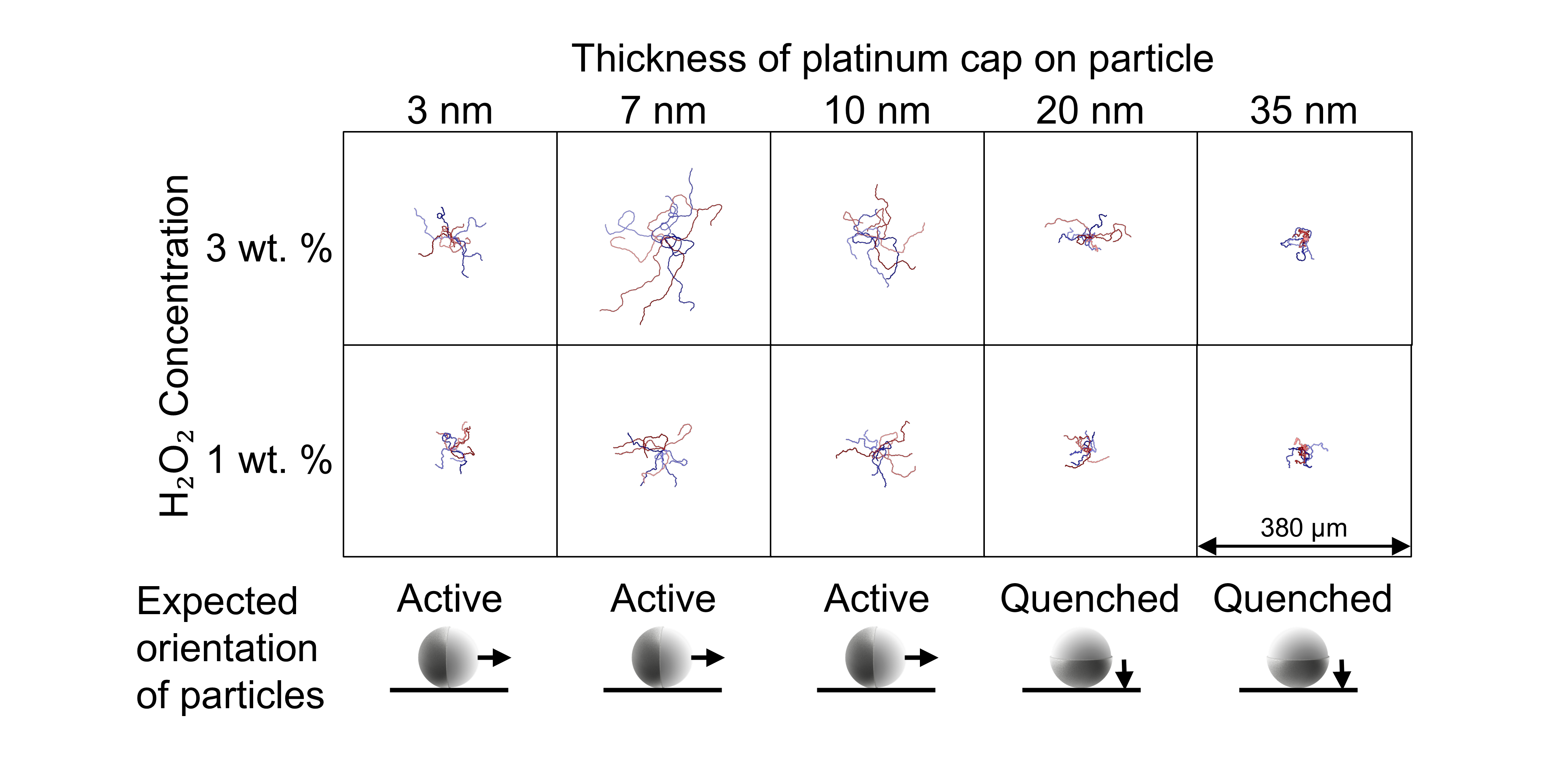}%
\caption{\label{FIG. 3.} 100 seconds trajectories of particles of different cap thickness in 1 wt./vol \% and 3 wt./vol \% hydrogen peroxide in water solution. Blue tracks show ceiling, while red tracks show flooring. 5 average velocity particles for ceiling and flooring in each condition are shown. Expected orientation for each thickness is shown below for each particle cap thickness.}
\end{figure*}
These studies suggest a critical coupling between particle orientation, the proximity and chemistry of the boundary, and propulsion speed. Herein, we aim to investigate how cap weight and boundary preparation couple to the apparent propulsion speed and the likelihood of a particle to move along either the top or bottom boundaries (i.e. ceiling or flooring). We fabricated polystyrene Janus particles with systematically varied platinum cap thickness, observed the location of propulsion (top vs. bottom boundary), measured their propulsion speed, and analyzed their trajectories.  These measurements revealed a compelling picture with newly discovered features, including how orientational quenching plays a critical role in propulsion dynamics and the influence of substrate preparation. In general, we found a strong coupling between these interfacial characteristics that are often neglected in experiments. We hope the research community finds these results useful in better understanding previous studies.

\textit{Experimental setup and methods} - Sulfate latex polystyrene particles (3 $\mu$m diameter, 8\% wt./vol) were purchased from Fisher Scientific (Lot \# 2855924). Particles were washed with ultra-pure water (18.2 M$\Omega$) five times before concentrating the particle solution to ~9\% wt./ vol. A monolayer of particles was formed on 20 mm × 20 mm silicon wafer using the method described in \cite{park_uniform_2019}. Platinum layers of 3 nm, 7nm, 10 nm, 20 nm, and 35 nm thickness were deposited onto the monolayers using Physical Vapor Deposition  (Angstrom Engineering Evovac Deposition System ). The deposition rate was maintained at 2 \AA/sec. Particles were removed from the wafer by sonication and washed with ultra-pure water five times. The particle solution was diluted to achieve ~10 particles per field of view (341 $\mu$m x 408 $\mu$m  under 20x objective). Particles were suspended in 30 \% wt./vol peroxide solution for 30 minutes for surface activation before measurement as described in \cite{campbell_gravitaxis_2013}. Particles were separated and washed after exposure to 30 \% hydrogen peroxide solution before adding peroxide to achieve the desired concentration of peroxide. Untreated glass slides were used from packaging as is, while plasma cleaned glass slides were sonicated in iso-propanol alcohol and acetone for 15 minutes each before air plasma treatment (Tergeo) for 2 minutes. The fluid cell was assembled by attaching secure seal spacer (diameter = 0.9 mm, height= 0.12 mm) onto the glass slide as in \cite{issa_engineered_2023}. The inner lining of the spacer was marked with a hydrophobic pen to avoid particle loss through pores. The spacer was closed by attaching a cover slip, which also served as the top boundary for the particles in the fluid cell. Particles were recorded at both the upper and lower planes immediately after assembling the fluid cell. Videos of length 100 seconds were recorded at 10 fps (ThorLabs camera) alternating between top and bottom boundaries for 20 minutes. If no particles were found at the top boundary, particles only at the bottom boundary were recorded and stored as .avi files. Videos were processed using ImageJ and in-house developed code to find mean-squared displacement, velocities, and the maximum displacement associated with a given trajectory. The detailed methodology is given in supplementary information of this letter.

\textit{Flooring and ceiling dynamics as a function of cap thickness} - Janus particles suspended in peroxide solution undergo a combination of stochastic and deterministic motion that, overtime, results in accumulation of particles at either the top (i.e. ceiling) or bottom (i.e. flooring) boundaries, as illustrated in Figure \ref{fig:setup}.  While theoretical studies have predicted the bifurcation of populations into flooring and ceiling states, few experimental studies have simultaneously observed both. Herein, we present measurements to gauge the importance of tracking both populations. Figure \ref{FIG. 2.} shows measurements of velocity and maximum displacement for both flooring and ceiling populations as a function of cap thickness. Interestingly, although there is a dependence on cap thickness (as discussed later), there is no measurable difference in velocity or displacement between the flooring and ceiling populations. For all experimental systems, from cap thicknesses of 3 nm to 35 nm, flooring and ceiling populations propel with statistically similar velocities and experience similar displacements. Note there remains lack of consensus in the community, in that Das et al. have argued that ceiling and flooring are similar, whereas Wang et al. presented data showing that the flooring velocity is higher than the ceiling velocity due to lower tilt on the floor\cite{das_floor-_2020, wang_tilt_2025}.

\begin{figure*}
\includegraphics[width=\linewidth]{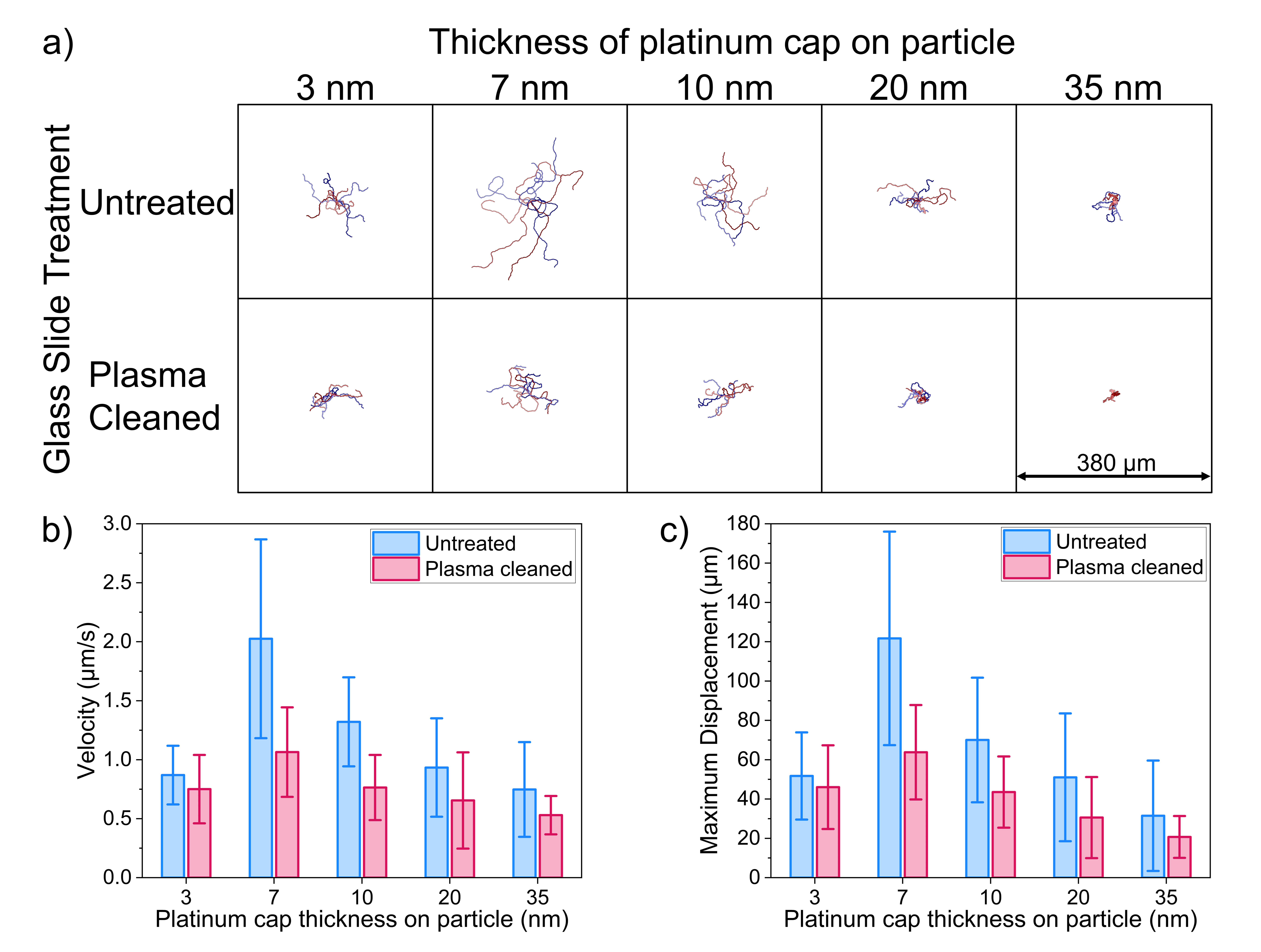}%
\caption{\label{FIG. 4.} a) Trajectories of particles of different cap thickness on untreated glass slide and plasma celaned glass slide. b) Velocities on untreated and plasma cleaned glass slide particles flooring in 3 wt./vol \% hydrogen peroxide in water solution. b) Maximum dispalcement on untreated and plasma cleaned glass slide particles flooring in 3 wt./vol \% hydrogen peroxide in water solution. Bars represent mean value with error bars showing standard deviation.}
\end{figure*}

Although there is not a strong relationship between velocity or displacement on the location of the Janus particle (ceiling vs flooring), there is a strong dependence of velocity on cap thickness. Figure \ref{FIG. 2.} shows that both the velocity and maximum displacement initially increased and then decreased for particles with caps of 3 nm, 7 nm, 10 nm, 20 nm, and 35 nm platinum cap thickness.  The initial increase in velocity from 3 nm to 7 nm is likely a consequence of non-uniform and insufficient platinum on the surface to react on as shown in previous studies for particles below 5 nm cap thickness \cite{meng_effect_2022}.  Meng et al. have shown that increasing platinum cap thickness from 1 nm to 5 nm leads to a rise in velocity, after which the velocity plateaus and remains unchanged up to 10 nm \cite{meng_effect_2022}. In our study, Janus particles with cap thicknesses of 7 nm and larger decreased in both velocity and maximum displacement with cap thickness. 

Trajectories were analyzed for Janus particles of systematic variation in cap thickness and at two different peroxide concentrations (see Figure \ref{FIG. 3.}). Particles with the same cap thickness traveled lesser in 1 wt. \% peroxide solution than 3 wt. \% peroxide solution. A similar trend can be seen in quantitative data such as velocity of particles and maximum displacement traveled by particles from initial position as shown in Figure \ref{FIG. 2.}. An increase in speed with increasing peroxide concentration agrees with previous literature as higher concentration of peroxide causes a higher reaction rate on the platinum surface, thus resulting in a higher propulsion force for particles \cite{howse_self-motile_2007}.  
\begin{table}[htbp]

\small % or \scriptsize
\centering
\caption{\label{Table 1}Ratio of ceiling  to flooring particles under different conditions. Red fill shows low, blue fill shows high.}
\renewcommand{\arraystretch}{1.1}
\resizebox{\columnwidth}{!}{  % resize to column width
\begin{tabular}{
  >{\centering\arraybackslash}m{2.4cm}  % Surface Treatment
  >{\centering\arraybackslash}m{1.4cm}  % H2O2
  >{\centering\arraybackslash}m{0.9cm}
  >{\centering\arraybackslash}m{0.9cm}
  >{\centering\arraybackslash}m{0.9cm}
  >{\centering\arraybackslash}m{0.9cm}
  >{\centering\arraybackslash}m{0.9cm}
  >{\centering\arraybackslash}m{0.8cm} % Legend
}
\toprule
\textbf{Surface Treatment} & \textbf{H\textsubscript{2}O\textsubscript{2} (\%)} 
  & \multicolumn{5}{c}{\textbf{Platinum Cap Thickness (nm)}} 
  & \multirow{3}{*}{\textbf{Legend}} \\
\cmidrule(lr){3-7}
{} & {} & \textbf{3} & \textbf{7} & \textbf{10} & \textbf{20} & \textbf{35} & \\
\midrule
\multirow{2}{*}{Untreated} 
  & 3\% 
    & \cellrb{0.45}{c45} & \cellrb{0.66}{c66} & \cellrb{0.75}{c75} & \cellrb{0.65}{c65} & \cellrb{0.36}{c36}
  & \multirow{4}{*}{\begin{tikzpicture}[x=0.6cm, y=1.2cm]
  \shade[top color=blue, bottom color=red] (0,0) rectangle (0.3,1);
  \draw (0,0) rectangle (0.3,1);
  \node[anchor=east] at (0,0) {\small 0.0};
  \node[anchor=east] at (0,0.5) {\small 0.5};
  \node[anchor=east] at (0,1) {\small 1.0};
\end{tikzpicture}
} \\
  & 1\% 
    & \cellrb{0.34}{c34} & \cellrb{0.66}{c66} & \cellrb{0.69}{c69} & \cellrb{0.14}{c14} & \cellrb{0.16}{c16} & \\
\cmidrule(lr){1-7}
\multirow{2}{*}{Plasma Cleaned} 
  & 3\% 
    & \cellrb{0.28}{c28} & \cellrb{0.82}{c82} & \cellrb{0.46}{c46} & \cellrb{0.46}{c46} & \cellrb{0.00}{c00} & \\
  & 1\% 
    & \cellrb{0.24}{c24} & \cellrb{0.74}{c74} & \cellrb{0.00}{c00} & \cellrb{0.00}{c00} & \cellrb{0.00}{c00} & \\
\bottomrule
\end{tabular}
}
\end{table}
Trajectories revealed a significant difference as a function of cap thickness with a trend similar to the velocity (see Figure \ref{FIG. 3.} and SI). The area distribution of trajectories, nominally considered to be the area covered by the ensemble of measurements, initially grew in size when increasing the cap thickness from 3 nm to 7 nm. However, once surpassing 7 nm in cap thickness, the area distribution of trajectories decreased from 7 nm and larger. This decrease in area distribution coincides with a decrease in maximum displacement. The reduction in area distribution of trajectories was not solely a consequence of reduction in speed as can be seen from the MSD (see SI). As characterized by the MSD, the trajectories transitioned from deterministically (MSD $\sim$ $tau^2$) to stochastically (MSD $\sim$ $tau$) dominated. This behavior is consistent with a re-orientation of the Janus particle such that the cap is pointing downward. This is due to the gravitational torque acting on the particles, thus pulling the orientation of cap towards the bottom boundary as shown in Figure \ref{FIG. 3.} in the column below. This becomes more evident as will be discussed below and summarized in Table \ref{Table 1}; the percentage of particles in the ceiling state also reduces with increasing cap thickness. 

\textit{Influence of surface treatment} - Plasma cleaning of the glass slide revealed that, for all platinum cap thicknesses, both ceiling and flooring particles migrated a shorter distance from the origin compared to those on untreated slides, as shown in Figure \ref{FIG. 4.}a. Quantitatively, the trend of flooring can be seen in Figure \ref{FIG. 4.} b and c,  where both the velocities and maximum displacement of particles decreased with increasing cap thickness except 3 nm thickness particles. Particles with caps of 3 nm thickness behaved in this fashion due to non-uniform platinum coverage driving the reaction, thus showing decreased activity. Particles with 7 nm cap thickness on untreated glass slide showed higher velocities of 2 $\mu$m/sec as compared to 1 $\mu$m/sec  for plasma cleaned glass slide. 

Perhaps most surprising is that this decrease in velocity is observed for both flooring and ceiling particles, when only the bottom surface had been plasma cleaned. Thus, the reduction in velocity cannot only be due to charge imparted to the bottom slide during plasma cleaning. Rather, we propose this effect results from a diffusion of charge into the solution from the bottom surface, which then slows down particles on both top and bottom boundaries. The effect is consistent across particles of different cap thickness, further confirming that charge released into solution slow down the particles. Furthermore, since ions in the solution slow down activity of particles, particles are more likely to go in quenched state in the absence of active propulsion force. This can be seen for 20 nm and 35 nm cap thickness particles on the plasma cleaned glass slide surface. Particles show almost no activity, as cap weight is high and gravitational torque pulls particle cap towards the boundary surface. Similarly ions in the solution mitigate the propulsive force that drives the particle cap away from boundary.

Several studies have shown that addition of salt in the solution slows down the speed of active particles due to charged ions interfering with electrical fields around particles \cite{brown_ionic_2014}. To the best of our knowledge, this is the first report showing that plasma cleaning one boundary will affect the speed of particles on the opposite boundary which is not plasma cleaned. Note, however, that one previous study found that dissolution of charges from a boundary can reduce the speed of rods near a boundary as well as in bulk \cite{Surface_charges_wang_2018}. We expect a similar effect in which negative charges developed on the plasma cleaned surface transport into the solution and slow down the active particles on both top and bottom boundaries.

When comparing the percentage of particles sliding along the top boundary as compared to the total number of particles in the solution, it was observed that lighter particles tended to be more likely to slide along the top boundary. As can be seen in Table  \ref{Table 1}, in 3\% peroxide solution on untreated glass slide, 7 nm, 10 nm, and 20 nm particles are more likely to be ceiling than flooring. Whereas, under the same condition, 35 nm particles are more likely to be flooring than ceiling. When peroxide concentration is reduced to 1\%, only 7 nm and 10 nm particles are more likely to be ceiling, while 20 nm and 35 nm particles are more likely to be flooring. When comparing this to a plasma cleaned glass slide for the bottom boundary, the chances of ceiling reduced further. Only 7 nm cap thickness particles are more likely to be ceiling than flooring in both 1\% and 3\% peroxide solutions. This decrease in ceiling on plasma cleaned surface can be attributed to the slowdown of particles, which in turn we proposed is an affect of charge introduced into the solution by the bottom boundary.

\textit{Conclusion} - In summary, this study provides a systematic experimental investigation of Janus particle dynamics as a function of platinum cap thickness, peroxide concentration, and substrate treatment. While theoretical studies have long predicted bifurcation into ceiling and flooring states, our results confirm that both populations exhibit statistically similar velocities and displacements across a wide range of cap thicknesses, thereby underscoring the importance of tracking both states simultaneously. We demonstrate that propulsion strongly depends on platinum cap thickness and peroxide concentration, with thinner caps ($<5 nm$) limited by incomplete catalytic coverage, intermediate caps ($\approx 7–10 nm$) yielding peak velocities and displacements, and thicker caps ($>20 nm$) showing reduced activity due to gravitational torque and reorientation effects. Plasma cleaning of substrates revealed an unexpected global suppression of particle velocities at both boundaries, consistent with charge diffusion from the treated surface into solution, thereby establishing a new link between substrate conditioning and active particle propulsion. Collectively, these findings not only extend experimental validation of ceiling and flooring states but also uncover how catalytic layer thickness, chemical environment, and boundary conditions couple to determine active particle trajectories. This work provides critical insight into how Janus particle motion can be tuned and controlled, with implications for the rational design of active colloids.

\begin{acknowledgments}
Grateful to Dr. Marola Issa and Dr Prateek Dwivedi  for their feedback. This work was supported by the National Science Foundation (NSF) award no. 2314405.
\end{acknowledgments}


%
% ****** End of file apstemplate.tex ******



\begin{thebibliography}{42}%
\makeatletter
\providecommand \@ifxundefined [1]{%
 \@ifx{#1\undefined}
}%
\providecommand \@ifnum [1]{%
 \ifnum #1\expandafter \@firstoftwo
 \else \expandafter \@secondoftwo
 \fi
}%
\providecommand \@ifx [1]{%
 \ifx #1\expandafter \@firstoftwo
 \else \expandafter \@secondoftwo
 \fi
}%
\providecommand \natexlab [1]{#1}%
\providecommand \enquote  [1]{``#1''}%
\providecommand \bibnamefont  [1]{#1}%
\providecommand \bibfnamefont [1]{#1}%
\providecommand \citenamefont [1]{#1}%
\providecommand \href@noop [0]{\@secondoftwo}%
\providecommand \href [0]{\begingroup \@sanitize@url \@href}%
\providecommand \@href[1]{\@@startlink{#1}\@@href}%
\providecommand \@@href[1]{\endgroup#1\@@endlink}%
\providecommand \@sanitize@url [0]{\catcode `\\12\catcode `\$12\catcode `\&12\catcode `\#12\catcode `\^12\catcode `\_12\catcode `\%12\relax}%
\providecommand \@@startlink[1]{}%
\providecommand \@@endlink[0]{}%
\providecommand \url  [0]{\begingroup\@sanitize@url \@url }%
\providecommand \@url [1]{\endgroup\@href {#1}{\urlprefix }}%
\providecommand \urlprefix  [0]{URL }%
\providecommand \Eprint [0]{\href }%
\providecommand \doibase [0]{https://doi.org/}%
\providecommand \selectlanguage [0]{\@gobble}%
\providecommand \bibinfo  [0]{\@secondoftwo}%
\providecommand \bibfield  [0]{\@secondoftwo}%
\providecommand \translation [1]{[#1]}%
\providecommand \BibitemOpen [0]{}%
\providecommand \bibitemStop [0]{}%
\providecommand \bibitemNoStop [0]{.\EOS\space}%
\providecommand \EOS [0]{\spacefactor3000\relax}%
\providecommand \BibitemShut  [1]{\csname bibitem#1\endcsname}%
\let\auto@bib@innerbib\@empty
%</preamble>
\bibitem [{\citenamefont {Ramaswamy}(2017)}]{ramaswamy_active_2017}%
  \BibitemOpen
  \bibfield  {author} {\bibinfo {author} {\bibfnamefont {S.}~\bibnamefont {Ramaswamy}},\ }\href {https://doi.org/10.1088/1742-5468/aa6bc5} {\bibfield  {journal} {\bibinfo  {journal} {J. Stat. Mech.}\ }\textbf {\bibinfo {volume} {2017}},\ \bibinfo {pages} {054002} (\bibinfo {year} {2017})},\ \bibinfo {note} {publisher: {IOP} Publishing and {SISSA}}\BibitemShut {NoStop}%
\bibitem [{\citenamefont {Marchetti}\ \emph {et~al.}(2013)\citenamefont {Marchetti}, \citenamefont {Joanny}, \citenamefont {Ramaswamy}, \citenamefont {Liverpool}, \citenamefont {Prost}, \citenamefont {Rao},\ and\ \citenamefont {Simha}}]{RevModPhys.85.1143}%
  \BibitemOpen
  \bibfield  {author} {\bibinfo {author} {\bibfnamefont {M.~C.}\ \bibnamefont {Marchetti}}, \bibinfo {author} {\bibfnamefont {J.~F.}\ \bibnamefont {Joanny}}, \bibinfo {author} {\bibfnamefont {S.}~\bibnamefont {Ramaswamy}}, \bibinfo {author} {\bibfnamefont {T.~B.}\ \bibnamefont {Liverpool}}, \bibinfo {author} {\bibfnamefont {J.}~\bibnamefont {Prost}}, \bibinfo {author} {\bibfnamefont {M.}~\bibnamefont {Rao}},\ and\ \bibinfo {author} {\bibfnamefont {R.~A.}\ \bibnamefont {Simha}},\ }\href {https://doi.org/10.1103/RevModPhys.85.1143} {\bibfield  {journal} {\bibinfo  {journal} {Rev. Mod. Phys.}\ }\textbf {\bibinfo {volume} {85}},\ \bibinfo {pages} {1143} (\bibinfo {year} {2013})}\BibitemShut {NoStop}%
\bibitem [{\citenamefont {Ma}\ \emph {et~al.}(2015)\citenamefont {Ma}, \citenamefont {Yang}, \citenamefont {Zhao},\ and\ \citenamefont {Wu}}]{ma_inducing_2015}%
  \BibitemOpen
  \bibfield  {author} {\bibinfo {author} {\bibfnamefont {F.}~\bibnamefont {Ma}}, \bibinfo {author} {\bibfnamefont {X.}~\bibnamefont {Yang}}, \bibinfo {author} {\bibfnamefont {H.}~\bibnamefont {Zhao}},\ and\ \bibinfo {author} {\bibfnamefont {N.}~\bibnamefont {Wu}},\ }\href {https://doi.org/10.1103/PhysRevLett.115.208302} {\bibfield  {journal} {\bibinfo  {journal} {Phys. Rev. Lett.}\ }\textbf {\bibinfo {volume} {115}},\ \bibinfo {pages} {208302} (\bibinfo {year} {2015})},\ \bibinfo {note} {publisher: American Physical Society}\BibitemShut {NoStop}%
\bibitem [{\citenamefont {Tottori}\ \emph {et~al.}(2012)\citenamefont {Tottori}, \citenamefont {Zhang}, \citenamefont {Qiu}, \citenamefont {Krawczyk}, \citenamefont {Franco-Obregón},\ and\ \citenamefont {Nelson}}]{Tottori_Magnetic_2012}%
  \BibitemOpen
  \bibfield  {author} {\bibinfo {author} {\bibfnamefont {S.}~\bibnamefont {Tottori}}, \bibinfo {author} {\bibfnamefont {L.}~\bibnamefont {Zhang}}, \bibinfo {author} {\bibfnamefont {F.}~\bibnamefont {Qiu}}, \bibinfo {author} {\bibfnamefont {K.~K.}\ \bibnamefont {Krawczyk}}, \bibinfo {author} {\bibfnamefont {A.}~\bibnamefont {Franco-Obregón}},\ and\ \bibinfo {author} {\bibfnamefont {B.~J.}\ \bibnamefont {Nelson}},\ }\href {https://doi.org/https://doi.org/10.1002/adma.201103818} {\bibfield  {journal} {\bibinfo  {journal} {Advanced Materials}\ }\textbf {\bibinfo {volume} {24}},\ \bibinfo {pages} {811} (\bibinfo {year} {2012})},\ \Eprint {https://arxiv.org/abs/https://advanced.onlinelibrary.wiley.com/doi/pdf/10.1002/adma.201103818} {https://advanced.onlinelibrary.wiley.com/doi/pdf/10.1002/adma.201103818} \BibitemShut {NoStop}%
\bibitem [{\citenamefont {Buttinoni}\ \emph {et~al.}(2012)\citenamefont {Buttinoni}, \citenamefont {Volpe}, \citenamefont {Kümmel}, \citenamefont {Volpe},\ and\ \citenamefont {Bechinger}}]{buttinoni_active_2012}%
  \BibitemOpen
  \bibfield  {author} {\bibinfo {author} {\bibfnamefont {I.}~\bibnamefont {Buttinoni}}, \bibinfo {author} {\bibfnamefont {G.}~\bibnamefont {Volpe}}, \bibinfo {author} {\bibfnamefont {F.}~\bibnamefont {Kümmel}}, \bibinfo {author} {\bibfnamefont {G.}~\bibnamefont {Volpe}},\ and\ \bibinfo {author} {\bibfnamefont {C.}~\bibnamefont {Bechinger}},\ }\href {https://doi.org/10.1088/0953-8984/24/28/284129} {\bibfield  {journal} {\bibinfo  {journal} {J. Phys.: Condens. Matter}\ }\textbf {\bibinfo {volume} {24}},\ \bibinfo {pages} {284129} (\bibinfo {year} {2012})},\ \bibinfo {note} {publisher: {IOP} Publishing}\BibitemShut {NoStop}%
\bibitem [{\citenamefont {Shields}\ and\ \citenamefont {Velev}(2017)}]{shields_evolution_2017}%
  \BibitemOpen
  \bibfield  {author} {\bibinfo {author} {\bibfnamefont {C.~W.}\ \bibnamefont {Shields}}\ and\ \bibinfo {author} {\bibfnamefont {O.~D.}\ \bibnamefont {Velev}},\ }\href {https://doi.org/10.1016/j.chempr.2017.09.006} {\bibfield  {journal} {\bibinfo  {journal} {Chem}\ }\textbf {\bibinfo {volume} {3}},\ \bibinfo {pages} {539} (\bibinfo {year} {2017})}\BibitemShut {NoStop}%
\bibitem [{\citenamefont {Peng}\ \emph {et~al.}(2020)\citenamefont {Peng}, \citenamefont {Chen}, \citenamefont {Kollipara}, \citenamefont {Liu}, \citenamefont {Fang}, \citenamefont {Lin},\ and\ \citenamefont {Zheng}}]{peng_opto-thermoelectric_2020}%
  \BibitemOpen
  \bibfield  {author} {\bibinfo {author} {\bibfnamefont {X.}~\bibnamefont {Peng}}, \bibinfo {author} {\bibfnamefont {Z.}~\bibnamefont {Chen}}, \bibinfo {author} {\bibfnamefont {P.~S.}\ \bibnamefont {Kollipara}}, \bibinfo {author} {\bibfnamefont {Y.}~\bibnamefont {Liu}}, \bibinfo {author} {\bibfnamefont {J.}~\bibnamefont {Fang}}, \bibinfo {author} {\bibfnamefont {L.}~\bibnamefont {Lin}},\ and\ \bibinfo {author} {\bibfnamefont {Y.}~\bibnamefont {Zheng}},\ }\href {https://doi.org/10.1038/s41377-020-00378-5} {\bibfield  {journal} {\bibinfo  {journal} {Light Sci Appl}\ }\textbf {\bibinfo {volume} {9}},\ \bibinfo {pages} {141} (\bibinfo {year} {2020})},\ \bibinfo {note} {publisher: Nature Publishing Group}\BibitemShut {NoStop}%
\bibitem [{\citenamefont {Simmchen}\ \emph {et~al.}(2016)\citenamefont {Simmchen}, \citenamefont {Katuri}, \citenamefont {Uspal}, \citenamefont {Popescu}, \citenamefont {Tasinkevych},\ and\ \citenamefont {Sánchez}}]{simmchen_topographical_2016}%
  \BibitemOpen
  \bibfield  {author} {\bibinfo {author} {\bibfnamefont {J.}~\bibnamefont {Simmchen}}, \bibinfo {author} {\bibfnamefont {J.}~\bibnamefont {Katuri}}, \bibinfo {author} {\bibfnamefont {W.~E.}\ \bibnamefont {Uspal}}, \bibinfo {author} {\bibfnamefont {M.~N.}\ \bibnamefont {Popescu}}, \bibinfo {author} {\bibfnamefont {M.}~\bibnamefont {Tasinkevych}},\ and\ \bibinfo {author} {\bibfnamefont {S.}~\bibnamefont {Sánchez}},\ }\href {https://doi.org/10.1038/ncomms10598} {\bibfield  {journal} {\bibinfo  {journal} {Nat Commun}\ }\textbf {\bibinfo {volume} {7}},\ \bibinfo {pages} {10598} (\bibinfo {year} {2016})}\BibitemShut {NoStop}%
\bibitem [{\citenamefont {Sundararajan}\ \emph {et~al.}(2008)\citenamefont {Sundararajan}, \citenamefont {Lammert}, \citenamefont {Zudans}, \citenamefont {Crespi},\ and\ \citenamefont {Sen}}]{sundararajan_catalytic_2008}%
  \BibitemOpen
  \bibfield  {author} {\bibinfo {author} {\bibfnamefont {S.}~\bibnamefont {Sundararajan}}, \bibinfo {author} {\bibfnamefont {P.~E.}\ \bibnamefont {Lammert}}, \bibinfo {author} {\bibfnamefont {A.~W.}\ \bibnamefont {Zudans}}, \bibinfo {author} {\bibfnamefont {V.~H.}\ \bibnamefont {Crespi}},\ and\ \bibinfo {author} {\bibfnamefont {A.}~\bibnamefont {Sen}},\ }\href {https://doi.org/10.1021/nl072275j} {\bibfield  {journal} {\bibinfo  {journal} {Nano Lett.}\ }\textbf {\bibinfo {volume} {8}},\ \bibinfo {pages} {1271} (\bibinfo {year} {2008})},\ \bibinfo {note} {publisher: American Chemical Society}\BibitemShut {NoStop}%
\bibitem [{\citenamefont {Baraban}\ \emph {et~al.}(2012)\citenamefont {Baraban}, \citenamefont {Tasinkevych}, \citenamefont {Popescu}, \citenamefont {Sanchez}, \citenamefont {Dietrich},\ and\ \citenamefont {Schmidt}}]{baraban_transport_2012}%
  \BibitemOpen
  \bibfield  {author} {\bibinfo {author} {\bibfnamefont {L.}~\bibnamefont {Baraban}}, \bibinfo {author} {\bibfnamefont {M.}~\bibnamefont {Tasinkevych}}, \bibinfo {author} {\bibfnamefont {M.~N.}\ \bibnamefont {Popescu}}, \bibinfo {author} {\bibfnamefont {S.}~\bibnamefont {Sanchez}}, \bibinfo {author} {\bibfnamefont {S.}~\bibnamefont {Dietrich}},\ and\ \bibinfo {author} {\bibfnamefont {O.~G.}\ \bibnamefont {Schmidt}},\ }\href {https://doi.org/10.1039/C1SM06512B} {\bibfield  {journal} {\bibinfo  {journal} {Soft Matter}\ }\textbf {\bibinfo {volume} {8}},\ \bibinfo {pages} {48} (\bibinfo {year} {2012})}\BibitemShut {NoStop}%
\bibitem [{\citenamefont {Balasubramanian}\ \emph {et~al.}(2011)\citenamefont {Balasubramanian}, \citenamefont {Kagan}, \citenamefont {Jack~Hu}, \citenamefont {Campuzano}, \citenamefont {Lobo-Castañon}, \citenamefont {Lim}, \citenamefont {Kang}, \citenamefont {Zimmerman}, \citenamefont {Zhang},\ and\ \citenamefont {Wang}}]{balasubramanian_micromachine-enabled_2011}%
  \BibitemOpen
  \bibfield  {author} {\bibinfo {author} {\bibfnamefont {S.}~\bibnamefont {Balasubramanian}}, \bibinfo {author} {\bibfnamefont {D.}~\bibnamefont {Kagan}}, \bibinfo {author} {\bibfnamefont {C.-M.}\ \bibnamefont {Jack~Hu}}, \bibinfo {author} {\bibfnamefont {S.}~\bibnamefont {Campuzano}}, \bibinfo {author} {\bibfnamefont {M.~J.}\ \bibnamefont {Lobo-Castañon}}, \bibinfo {author} {\bibfnamefont {N.}~\bibnamefont {Lim}}, \bibinfo {author} {\bibfnamefont {D.~Y.}\ \bibnamefont {Kang}}, \bibinfo {author} {\bibfnamefont {M.}~\bibnamefont {Zimmerman}}, \bibinfo {author} {\bibfnamefont {L.}~\bibnamefont {Zhang}},\ and\ \bibinfo {author} {\bibfnamefont {J.}~\bibnamefont {Wang}},\ }\href {https://doi.org/10.1002/anie.201100115} {\bibfield  {journal} {\bibinfo  {journal} {Angewandte Chemie International Edition}\ }\textbf {\bibinfo {volume} {50}},\ \bibinfo {pages} {4161} (\bibinfo {year} {2011})},\ \bibinfo {note} {\_eprint: https://onlinelibrary.wiley.com/doi/pdf/10.1002/anie.201100115}\BibitemShut {NoStop}%
\bibitem [{\citenamefont {Xi}\ \emph {et~al.}(2013)\citenamefont {Xi}, \citenamefont {Solovev}, \citenamefont {Ananth}, \citenamefont {Gracias}, \citenamefont {Sanchez},\ and\ \citenamefont {Schmidt}}]{xi_rolled-up_2013}%
  \BibitemOpen
  \bibfield  {author} {\bibinfo {author} {\bibfnamefont {W.}~\bibnamefont {Xi}}, \bibinfo {author} {\bibfnamefont {A.~A.}\ \bibnamefont {Solovev}}, \bibinfo {author} {\bibfnamefont {A.~N.}\ \bibnamefont {Ananth}}, \bibinfo {author} {\bibfnamefont {D.~H.}\ \bibnamefont {Gracias}}, \bibinfo {author} {\bibfnamefont {S.}~\bibnamefont {Sanchez}},\ and\ \bibinfo {author} {\bibfnamefont {O.~G.}\ \bibnamefont {Schmidt}},\ }\href {https://doi.org/10.1039/C2NR32798H} {\bibfield  {journal} {\bibinfo  {journal} {Nanoscale}\ }\textbf {\bibinfo {volume} {5}},\ \bibinfo {pages} {1294} (\bibinfo {year} {2013})}\BibitemShut {NoStop}%
\bibitem [{\citenamefont {Bunea}\ \emph {et~al.}(2015)\citenamefont {Bunea}, \citenamefont {Pavel}, \citenamefont {David},\ and\ \citenamefont {Gáspár}}]{bunea_sensing_2015}%
  \BibitemOpen
  \bibfield  {author} {\bibinfo {author} {\bibfnamefont {A.-I.}\ \bibnamefont {Bunea}}, \bibinfo {author} {\bibfnamefont {I.-A.}\ \bibnamefont {Pavel}}, \bibinfo {author} {\bibfnamefont {S.}~\bibnamefont {David}},\ and\ \bibinfo {author} {\bibfnamefont {S.}~\bibnamefont {Gáspár}},\ }\href {https://doi.org/10.1016/j.bios.2014.05.062} {\bibfield  {journal} {\bibinfo  {journal} {Biosensors and Bioelectronics}\ }\bibinfo {series} {Special Issue: {BIOSENSORS} 2014},\ \textbf {\bibinfo {volume} {67}},\ \bibinfo {pages} {42} (\bibinfo {year} {2015})}\BibitemShut {NoStop}%
\bibitem [{\citenamefont {Soler}\ \emph {et~al.}(2013)\citenamefont {Soler}, \citenamefont {Magdanz}, \citenamefont {Fomin}, \citenamefont {Sanchez},\ and\ \citenamefont {Schmidt}}]{soler_self-propelled_2013}%
  \BibitemOpen
  \bibfield  {author} {\bibinfo {author} {\bibfnamefont {L.}~\bibnamefont {Soler}}, \bibinfo {author} {\bibfnamefont {V.}~\bibnamefont {Magdanz}}, \bibinfo {author} {\bibfnamefont {V.~M.}\ \bibnamefont {Fomin}}, \bibinfo {author} {\bibfnamefont {S.}~\bibnamefont {Sanchez}},\ and\ \bibinfo {author} {\bibfnamefont {O.~G.}\ \bibnamefont {Schmidt}},\ }\href {https://doi.org/10.1021/nn405075d} {\bibfield  {journal} {\bibinfo  {journal} {{ACS} Nano}\ }\textbf {\bibinfo {volume} {7}},\ \bibinfo {pages} {9611} (\bibinfo {year} {2013})},\ \bibinfo {note} {publisher: American Chemical Society}\BibitemShut {NoStop}%
\bibitem [{\citenamefont {Tohidi}\ \emph {et~al.}(2022)\citenamefont {Tohidi}, \citenamefont {Teimouri}, \citenamefont {Jafari}, \citenamefont {Gharibshahi},\ and\ \citenamefont {Omidkhah}}]{tohidi_application_2022}%
  \BibitemOpen
  \bibfield  {author} {\bibinfo {author} {\bibfnamefont {Z.}~\bibnamefont {Tohidi}}, \bibinfo {author} {\bibfnamefont {A.}~\bibnamefont {Teimouri}}, \bibinfo {author} {\bibfnamefont {A.}~\bibnamefont {Jafari}}, \bibinfo {author} {\bibfnamefont {R.}~\bibnamefont {Gharibshahi}},\ and\ \bibinfo {author} {\bibfnamefont {M.~R.}\ \bibnamefont {Omidkhah}},\ }\href {https://doi.org/10.1016/j.petrol.2021.109602} {\bibfield  {journal} {\bibinfo  {journal} {Journal of Petroleum Science and Engineering}\ }\textbf {\bibinfo {volume} {208}},\ \bibinfo {pages} {109602} (\bibinfo {year} {2022})}\BibitemShut {NoStop}%
\bibitem [{\citenamefont {Xu}\ \emph {et~al.}(2018)\citenamefont {Xu}, \citenamefont {Medina-Sánchez}, \citenamefont {Magdanz}, \citenamefont {Schwarz}, \citenamefont {Hebenstreit},\ and\ \citenamefont {Schmidt}}]{xu_sperm-hybrid_2018}%
  \BibitemOpen
  \bibfield  {author} {\bibinfo {author} {\bibfnamefont {H.}~\bibnamefont {Xu}}, \bibinfo {author} {\bibfnamefont {M.}~\bibnamefont {Medina-Sánchez}}, \bibinfo {author} {\bibfnamefont {V.}~\bibnamefont {Magdanz}}, \bibinfo {author} {\bibfnamefont {L.}~\bibnamefont {Schwarz}}, \bibinfo {author} {\bibfnamefont {F.}~\bibnamefont {Hebenstreit}},\ and\ \bibinfo {author} {\bibfnamefont {O.~G.}\ \bibnamefont {Schmidt}},\ }\href {https://doi.org/10.1021/acsnano.7b06398} {\bibfield  {journal} {\bibinfo  {journal} {{ACS} Nano}\ }\textbf {\bibinfo {volume} {12}},\ \bibinfo {pages} {327} (\bibinfo {year} {2018})},\ \bibinfo {note} {publisher: American Chemical Society}\BibitemShut {NoStop}%
\bibitem [{\citenamefont {Nsamela}\ \emph {et~al.}(2023)\citenamefont {Nsamela}, \citenamefont {Garcia~Zintzun}, \citenamefont {Montenegro-Johnson},\ and\ \citenamefont {Simmchen}}]{nsamela_colloidal_2023}%
  \BibitemOpen
  \bibfield  {author} {\bibinfo {author} {\bibfnamefont {A.}~\bibnamefont {Nsamela}}, \bibinfo {author} {\bibfnamefont {A.~I.}\ \bibnamefont {Garcia~Zintzun}}, \bibinfo {author} {\bibfnamefont {T.~D.}\ \bibnamefont {Montenegro-Johnson}},\ and\ \bibinfo {author} {\bibfnamefont {J.}~\bibnamefont {Simmchen}},\ }\href {https://doi.org/10.1002/smll.202202685} {\bibfield  {journal} {\bibinfo  {journal} {Small}\ }\textbf {\bibinfo {volume} {19}},\ \bibinfo {pages} {2202685} (\bibinfo {year} {2023})},\ \bibinfo {note} {\_eprint: https://onlinelibrary.wiley.com/doi/pdf/10.1002/smll.202202685}\BibitemShut {NoStop}%
\bibitem [{\citenamefont {Ramia}\ \emph {et~al.}(1993)\citenamefont {Ramia}, \citenamefont {Tullock},\ and\ \citenamefont {Phan-Thien}}]{ramia_role_1993}%
  \BibitemOpen
  \bibfield  {author} {\bibinfo {author} {\bibfnamefont {M.}~\bibnamefont {Ramia}}, \bibinfo {author} {\bibfnamefont {D.~L.}\ \bibnamefont {Tullock}},\ and\ \bibinfo {author} {\bibfnamefont {N.}~\bibnamefont {Phan-Thien}},\ }\href {https://doi.org/10.1016/S0006-3495(93)81129-9} {\bibfield  {journal} {\bibinfo  {journal} {Biophysical Journal}\ }\textbf {\bibinfo {volume} {65}},\ \bibinfo {pages} {755} (\bibinfo {year} {1993})},\ \bibinfo {note} {publisher: Elsevier}\BibitemShut {NoStop}%
\bibitem [{\citenamefont {Howse}\ \emph {et~al.}(2007)\citenamefont {Howse}, \citenamefont {Jones}, \citenamefont {Ryan}, \citenamefont {Gough}, \citenamefont {Vafabakhsh},\ and\ \citenamefont {Golestanian}}]{howse_self-motile_2007}%
  \BibitemOpen
  \bibfield  {author} {\bibinfo {author} {\bibfnamefont {J.~R.}\ \bibnamefont {Howse}}, \bibinfo {author} {\bibfnamefont {R.~A.~L.}\ \bibnamefont {Jones}}, \bibinfo {author} {\bibfnamefont {A.~J.}\ \bibnamefont {Ryan}}, \bibinfo {author} {\bibfnamefont {T.}~\bibnamefont {Gough}}, \bibinfo {author} {\bibfnamefont {R.}~\bibnamefont {Vafabakhsh}},\ and\ \bibinfo {author} {\bibfnamefont {R.}~\bibnamefont {Golestanian}},\ }\href {https://doi.org/10.1103/PhysRevLett.99.048102} {\bibfield  {journal} {\bibinfo  {journal} {Phys. Rev. Lett.}\ }\textbf {\bibinfo {volume} {99}},\ \bibinfo {pages} {048102} (\bibinfo {year} {2007})}\BibitemShut {NoStop}%
\bibitem [{\citenamefont {Ebbens}\ \emph {et~al.}(2012)\citenamefont {Ebbens}, \citenamefont {Tu}, \citenamefont {Howse},\ and\ \citenamefont {Golestanian}}]{ebbens_size_2012}%
  \BibitemOpen
  \bibfield  {author} {\bibinfo {author} {\bibfnamefont {S.}~\bibnamefont {Ebbens}}, \bibinfo {author} {\bibfnamefont {M.-H.}\ \bibnamefont {Tu}}, \bibinfo {author} {\bibfnamefont {J.~R.}\ \bibnamefont {Howse}},\ and\ \bibinfo {author} {\bibfnamefont {R.}~\bibnamefont {Golestanian}},\ }\href {https://doi.org/10.1103/PhysRevE.85.020401} {\bibfield  {journal} {\bibinfo  {journal} {Phys. Rev. E}\ }\textbf {\bibinfo {volume} {85}},\ \bibinfo {pages} {020401} (\bibinfo {year} {2012})},\ \bibinfo {note} {publisher: American Physical Society}\BibitemShut {NoStop}%
\bibitem [{\citenamefont {Das}\ \emph {et~al.}(2020)\citenamefont {Das}, \citenamefont {Jalilvand}, \citenamefont {Popescu}, \citenamefont {Uspal}, \citenamefont {Dietrich},\ and\ \citenamefont {Kretzschmar}}]{das_floor-_2020}%
  \BibitemOpen
  \bibfield  {author} {\bibinfo {author} {\bibfnamefont {S.}~\bibnamefont {Das}}, \bibinfo {author} {\bibfnamefont {Z.}~\bibnamefont {Jalilvand}}, \bibinfo {author} {\bibfnamefont {M.~N.}\ \bibnamefont {Popescu}}, \bibinfo {author} {\bibfnamefont {W.~E.}\ \bibnamefont {Uspal}}, \bibinfo {author} {\bibfnamefont {S.}~\bibnamefont {Dietrich}},\ and\ \bibinfo {author} {\bibfnamefont {I.}~\bibnamefont {Kretzschmar}},\ }\href {https://doi.org/10.1021/acs.langmuir.9b03696} {\bibfield  {journal} {\bibinfo  {journal} {Langmuir}\ }\textbf {\bibinfo {volume} {36}},\ \bibinfo {pages} {7133} (\bibinfo {year} {2020})},\ \bibinfo {note} {publisher: American Chemical Society}\BibitemShut {NoStop}%
\bibitem [{\citenamefont {Brown}\ and\ \citenamefont {Poon}(2014)}]{brown_ionic_2014}%
  \BibitemOpen
  \bibfield  {author} {\bibinfo {author} {\bibfnamefont {A.}~\bibnamefont {Brown}}\ and\ \bibinfo {author} {\bibfnamefont {W.}~\bibnamefont {Poon}},\ }\href {https://doi.org/10.1039/C4SM00340C} {\bibfield  {journal} {\bibinfo  {journal} {Soft Matter}\ }\textbf {\bibinfo {volume} {10}},\ \bibinfo {pages} {4016} (\bibinfo {year} {2014})}\BibitemShut {NoStop}%
\bibitem [{\citenamefont {Wang}(2023)}]{wang_open_2023}%
  \BibitemOpen
  \bibfield  {author} {\bibinfo {author} {\bibfnamefont {W.}~\bibnamefont {Wang}},\ }\href {https://doi.org/10.1021/jacs.3c09223} {\bibfield  {journal} {\bibinfo  {journal} {J. Am. Chem. Soc.}\ }\textbf {\bibinfo {volume} {145}},\ \bibinfo {pages} {27185} (\bibinfo {year} {2023})}\BibitemShut {NoStop}%
\bibitem [{\citenamefont {Das}\ \emph {et~al.}(2015)\citenamefont {Das}, \citenamefont {Garg}, \citenamefont {Campbell}, \citenamefont {Howse}, \citenamefont {Sen}, \citenamefont {Velegol}, \citenamefont {Golestanian},\ and\ \citenamefont {Ebbens}}]{das_boundaries_2015}%
  \BibitemOpen
  \bibfield  {author} {\bibinfo {author} {\bibfnamefont {S.}~\bibnamefont {Das}}, \bibinfo {author} {\bibfnamefont {A.}~\bibnamefont {Garg}}, \bibinfo {author} {\bibfnamefont {A.~I.}\ \bibnamefont {Campbell}}, \bibinfo {author} {\bibfnamefont {J.}~\bibnamefont {Howse}}, \bibinfo {author} {\bibfnamefont {A.}~\bibnamefont {Sen}}, \bibinfo {author} {\bibfnamefont {D.}~\bibnamefont {Velegol}}, \bibinfo {author} {\bibfnamefont {R.}~\bibnamefont {Golestanian}},\ and\ \bibinfo {author} {\bibfnamefont {S.~J.}\ \bibnamefont {Ebbens}},\ }\href {https://doi.org/10.1038/ncomms9999} {\bibfield  {journal} {\bibinfo  {journal} {Nat Commun}\ }\textbf {\bibinfo {volume} {6}},\ \bibinfo {pages} {8999} (\bibinfo {year} {2015})}\BibitemShut {NoStop}%
\bibitem [{\citenamefont {Mozaffari}\ \emph {et~al.}(2016)\citenamefont {Mozaffari}, \citenamefont {Sharifi-Mood}, \citenamefont {Koplik},\ and\ \citenamefont {Maldarelli}}]{mozaffari_self-diffusiophoretic_2016}%
  \BibitemOpen
  \bibfield  {author} {\bibinfo {author} {\bibfnamefont {A.}~\bibnamefont {Mozaffari}}, \bibinfo {author} {\bibfnamefont {N.}~\bibnamefont {Sharifi-Mood}}, \bibinfo {author} {\bibfnamefont {J.}~\bibnamefont {Koplik}},\ and\ \bibinfo {author} {\bibfnamefont {C.}~\bibnamefont {Maldarelli}},\ }\href {https://doi.org/10.1063/1.4948398} {\bibfield  {journal} {\bibinfo  {journal} {Physics of Fluids}\ }\textbf {\bibinfo {volume} {28}},\ \bibinfo {pages} {053107} (\bibinfo {year} {2016})}\BibitemShut {NoStop}%
\bibitem [{\citenamefont {Uspal}\ \emph {et~al.}(2014)\citenamefont {Uspal}, \citenamefont {Popescu}, \citenamefont {Dietrich},\ and\ \citenamefont {Tasinkevych}}]{uspal_self-propulsion_2014}%
  \BibitemOpen
  \bibfield  {author} {\bibinfo {author} {\bibfnamefont {W.~E.}\ \bibnamefont {Uspal}}, \bibinfo {author} {\bibfnamefont {M.~N.}\ \bibnamefont {Popescu}}, \bibinfo {author} {\bibfnamefont {S.}~\bibnamefont {Dietrich}},\ and\ \bibinfo {author} {\bibfnamefont {M.}~\bibnamefont {Tasinkevych}},\ }\href {https://doi.org/10.1039/C4SM02317J} {\bibfield  {journal} {\bibinfo  {journal} {Soft Matter}\ }\textbf {\bibinfo {volume} {11}},\ \bibinfo {pages} {434} (\bibinfo {year} {2014})},\ \bibinfo {note} {publisher: The Royal Society of Chemistry}\BibitemShut {NoStop}%
\bibitem [{\citenamefont {Xiao}\ \emph {et~al.}(2020)\citenamefont {Xiao}, \citenamefont {Duan}, \citenamefont {Xu}, \citenamefont {Cui}, \citenamefont {Zhang},\ and\ \citenamefont {Wang}}]{xiao_synergistic_2020}%
  \BibitemOpen
  \bibfield  {author} {\bibinfo {author} {\bibfnamefont {Z.}~\bibnamefont {Xiao}}, \bibinfo {author} {\bibfnamefont {S.}~\bibnamefont {Duan}}, \bibinfo {author} {\bibfnamefont {P.}~\bibnamefont {Xu}}, \bibinfo {author} {\bibfnamefont {J.}~\bibnamefont {Cui}}, \bibinfo {author} {\bibfnamefont {H.}~\bibnamefont {Zhang}},\ and\ \bibinfo {author} {\bibfnamefont {W.}~\bibnamefont {Wang}},\ }\href {https://doi.org/10.1021/acsnano.0c03022} {\bibfield  {journal} {\bibinfo  {journal} {{ACS} Nano}\ }\textbf {\bibinfo {volume} {14}},\ \bibinfo {pages} {8658} (\bibinfo {year} {2020})}\BibitemShut {NoStop}%
\bibitem [{\citenamefont {Ketzetzi}\ \emph {et~al.}(2020)\citenamefont {Ketzetzi}, \citenamefont {de~Graaf}, \citenamefont {Doherty},\ and\ \citenamefont {Kraft}}]{Slip_length_Ketzetzi_2020}%
  \BibitemOpen
  \bibfield  {author} {\bibinfo {author} {\bibfnamefont {S.}~\bibnamefont {Ketzetzi}}, \bibinfo {author} {\bibfnamefont {J.}~\bibnamefont {de~Graaf}}, \bibinfo {author} {\bibfnamefont {R.~P.}\ \bibnamefont {Doherty}},\ and\ \bibinfo {author} {\bibfnamefont {D.~J.}\ \bibnamefont {Kraft}},\ }\href {https://doi.org/10.1103/PhysRevLett.124.048002} {\bibfield  {journal} {\bibinfo  {journal} {Phys. Rev. Lett.}\ }\textbf {\bibinfo {volume} {124}},\ \bibinfo {pages} {048002} (\bibinfo {year} {2020})}\BibitemShut {NoStop}%
\bibitem [{\citenamefont {Rashidi}\ \emph {et~al.}(2020)\citenamefont {Rashidi}, \citenamefont {Razavi},\ and\ \citenamefont {Wirth}}]{rashidi_influence_2020}%
  \BibitemOpen
  \bibfield  {author} {\bibinfo {author} {\bibfnamefont {A.}~\bibnamefont {Rashidi}}, \bibinfo {author} {\bibfnamefont {S.}~\bibnamefont {Razavi}},\ and\ \bibinfo {author} {\bibfnamefont {C.~L.}\ \bibnamefont {Wirth}},\ }\href {https://doi.org/10.1103/PhysRevE.101.042606} {\bibfield  {journal} {\bibinfo  {journal} {Phys. Rev. E}\ }\textbf {\bibinfo {volume} {101}},\ \bibinfo {pages} {042606} (\bibinfo {year} {2020})}\BibitemShut {NoStop}%
\bibitem [{\citenamefont {Rashidi}\ and\ \citenamefont {Wirth}(2017)}]{rashidi_motion_2017}%
  \BibitemOpen
  \bibfield  {author} {\bibinfo {author} {\bibfnamefont {A.}~\bibnamefont {Rashidi}}\ and\ \bibinfo {author} {\bibfnamefont {C.~L.}\ \bibnamefont {Wirth}},\ }\href {https://doi.org/10.1063/1.4994843} {\bibfield  {journal} {\bibinfo  {journal} {The Journal of Chemical Physics}\ }\textbf {\bibinfo {volume} {147}},\ \bibinfo {pages} {224906} (\bibinfo {year} {2017})}\BibitemShut {NoStop}%
\bibitem [{\citenamefont {Rajupet}\ \emph {et~al.}(2021)\citenamefont {Rajupet}, \citenamefont {Rashidi},\ and\ \citenamefont {Wirth}}]{rajupet_derjaguin-landau-verwey-overbeek_2021}%
  \BibitemOpen
  \bibfield  {author} {\bibinfo {author} {\bibfnamefont {S.}~\bibnamefont {Rajupet}}, \bibinfo {author} {\bibfnamefont {A.}~\bibnamefont {Rashidi}},\ and\ \bibinfo {author} {\bibfnamefont {C.~L.}\ \bibnamefont {Wirth}},\ }\href {https://doi.org/10.1103/PhysRevE.103.032610} {\bibfield  {journal} {\bibinfo  {journal} {Phys. Rev. E}\ }\textbf {\bibinfo {volume} {103}},\ \bibinfo {pages} {032610} (\bibinfo {year} {2021})},\ \bibinfo {note} {publisher: American Physical Society}\BibitemShut {NoStop}%
\bibitem [{\citenamefont {Jalilvand}\ \emph {et~al.}(2025)\citenamefont {Jalilvand}, \citenamefont {Notarmuzi}, \citenamefont {Córdova-Figueroa}, \citenamefont {Bianchi},\ and\ \citenamefont {Kretzschmar}}]{bottom-heavy-dynamics-Jalilvand-2025}%
  \BibitemOpen
  \bibfield  {author} {\bibinfo {author} {\bibfnamefont {Z.}~\bibnamefont {Jalilvand}}, \bibinfo {author} {\bibfnamefont {D.}~\bibnamefont {Notarmuzi}}, \bibinfo {author} {\bibfnamefont {U.~M.}\ \bibnamefont {Córdova-Figueroa}}, \bibinfo {author} {\bibfnamefont {E.}~\bibnamefont {Bianchi}},\ and\ \bibinfo {author} {\bibfnamefont {I.}~\bibnamefont {Kretzschmar}},\ }\href {https://doi.org/10.1039/D5SM00229J} {\bibfield  {journal} {\bibinfo  {journal} {Soft Matter}\ }\textbf {\bibinfo {volume} {21}},\ \bibinfo {pages} {5773} (\bibinfo {year} {2025})}\BibitemShut {NoStop}%
\bibitem [{\citenamefont {Meng}\ \emph {et~al.}(2022)\citenamefont {Meng}, \citenamefont {Zhang}, \citenamefont {Liu}, \citenamefont {Zhang}, \citenamefont {Ma}, \citenamefont {Chen}, \citenamefont {Gao}, \citenamefont {Ma}, \citenamefont {Wang},\ and\ \citenamefont {Feng}}]{meng_effect_2022}%
  \BibitemOpen
  \bibfield  {author} {\bibinfo {author} {\bibfnamefont {S.}~\bibnamefont {Meng}}, \bibinfo {author} {\bibfnamefont {Y.}~\bibnamefont {Zhang}}, \bibinfo {author} {\bibfnamefont {Y.}~\bibnamefont {Liu}}, \bibinfo {author} {\bibfnamefont {Z.}~\bibnamefont {Zhang}}, \bibinfo {author} {\bibfnamefont {K.}~\bibnamefont {Ma}}, \bibinfo {author} {\bibfnamefont {X.}~\bibnamefont {Chen}}, \bibinfo {author} {\bibfnamefont {Q.}~\bibnamefont {Gao}}, \bibinfo {author} {\bibfnamefont {X.}~\bibnamefont {Ma}}, \bibinfo {author} {\bibfnamefont {W.}~\bibnamefont {Wang}},\ and\ \bibinfo {author} {\bibfnamefont {H.}~\bibnamefont {Feng}},\ }\href {https://doi.org/10.1016/j.jciso.2022.100046} {\bibfield  {journal} {\bibinfo  {journal} {{JCIS} Open}\ }\textbf {\bibinfo {volume} {5}},\ \bibinfo {pages} {100046} (\bibinfo {year} {2022})}\BibitemShut {NoStop}%
\bibitem [{\citenamefont {Rashidi}\ \emph {et~al.}(2018)\citenamefont {Rashidi}, \citenamefont {Issa}, \citenamefont {Martin}, \citenamefont {Avishai}, \citenamefont {Razavi},\ and\ \citenamefont {Wirth}}]{rashidi_local_2018}%
  \BibitemOpen
  \bibfield  {author} {\bibinfo {author} {\bibfnamefont {A.}~\bibnamefont {Rashidi}}, \bibinfo {author} {\bibfnamefont {M.~W.}\ \bibnamefont {Issa}}, \bibinfo {author} {\bibfnamefont {I.~T.}\ \bibnamefont {Martin}}, \bibinfo {author} {\bibfnamefont {A.}~\bibnamefont {Avishai}}, \bibinfo {author} {\bibfnamefont {S.}~\bibnamefont {Razavi}},\ and\ \bibinfo {author} {\bibfnamefont {C.~L.}\ \bibnamefont {Wirth}},\ }\href {https://doi.org/10.1021/acsami.8b11011} {\bibfield  {journal} {\bibinfo  {journal} {{ACS} Appl. Mater. Interfaces}\ }\textbf {\bibinfo {volume} {10}},\ \bibinfo {pages} {30925} (\bibinfo {year} {2018})}\BibitemShut {NoStop}%
\bibitem [{\citenamefont {Issa}\ \emph {et~al.}(2023)\citenamefont {Issa}, \citenamefont {Calderon}, \citenamefont {Kamlet}, \citenamefont {Asaei}, \citenamefont {Renner},\ and\ \citenamefont {Wirth}}]{issa_engineered_2023}%
  \BibitemOpen
  \bibfield  {author} {\bibinfo {author} {\bibfnamefont {M.~W.}\ \bibnamefont {Issa}}, \bibinfo {author} {\bibfnamefont {D.}~\bibnamefont {Calderon}}, \bibinfo {author} {\bibfnamefont {O.}~\bibnamefont {Kamlet}}, \bibinfo {author} {\bibfnamefont {S.}~\bibnamefont {Asaei}}, \bibinfo {author} {\bibfnamefont {J.~N.}\ \bibnamefont {Renner}},\ and\ \bibinfo {author} {\bibfnamefont {C.~L.}\ \bibnamefont {Wirth}},\ }\href {https://doi.org/10.1021/acsaenm.3c00263} {\bibfield  {journal} {\bibinfo  {journal} {{ACS} Appl. Eng. Mater.}\ }\textbf {\bibinfo {volume} {1}},\ \bibinfo {pages} {1983} (\bibinfo {year} {2023})},\ \bibinfo {note} {publisher: American Chemical Society}\BibitemShut {NoStop}%
\bibitem [{\citenamefont {Kalil}\ \emph {et~al.}(2021)\citenamefont {Kalil}, \citenamefont {Baumgartner}, \citenamefont {Issa}, \citenamefont {Ryan},\ and\ \citenamefont {Wirth}}]{kalil_influence_2021}%
  \BibitemOpen
  \bibfield  {author} {\bibinfo {author} {\bibfnamefont {M.~A.}\ \bibnamefont {Kalil}}, \bibinfo {author} {\bibfnamefont {N.~R.}\ \bibnamefont {Baumgartner}}, \bibinfo {author} {\bibfnamefont {M.~W.}\ \bibnamefont {Issa}}, \bibinfo {author} {\bibfnamefont {S.~D.}\ \bibnamefont {Ryan}},\ and\ \bibinfo {author} {\bibfnamefont {C.~L.}\ \bibnamefont {Wirth}},\ }\href {https://doi.org/10.1016/j.colsurfa.2021.127191} {\bibfield  {journal} {\bibinfo  {journal} {Colloids and Surfaces A: Physicochemical and Engineering Aspects}\ }\textbf {\bibinfo {volume} {627}},\ \bibinfo {pages} {127191} (\bibinfo {year} {2021})}\BibitemShut {NoStop}%
\bibitem [{\citenamefont {Pawar}\ and\ \citenamefont {Kretzschmar}(2009)}]{pawar2009multifunctional}%
  \BibitemOpen
  \bibfield  {author} {\bibinfo {author} {\bibfnamefont {A.~B.}\ \bibnamefont {Pawar}}\ and\ \bibinfo {author} {\bibfnamefont {I.}~\bibnamefont {Kretzschmar}},\ }\href@noop {} {\bibfield  {journal} {\bibinfo  {journal} {Langmuir}\ }\textbf {\bibinfo {volume} {25}},\ \bibinfo {pages} {9057} (\bibinfo {year} {2009})}\BibitemShut {NoStop}%
\bibitem [{\citenamefont {Liu}\ \emph {et~al.}(2025)\citenamefont {Liu}, \citenamefont {Xu}, \citenamefont {Qiao}, \citenamefont {Li}, \citenamefont {Ma}, \citenamefont {Kuang}, \citenamefont {Zhang},\ and\ \citenamefont {Wang}}]{wang_tilt_2025}%
  \BibitemOpen
  \bibfield  {author} {\bibinfo {author} {\bibfnamefont {J.}~\bibnamefont {Liu}}, \bibinfo {author} {\bibfnamefont {Y.}~\bibnamefont {Xu}}, \bibinfo {author} {\bibfnamefont {Z.}~\bibnamefont {Qiao}}, \bibinfo {author} {\bibfnamefont {S.}~\bibnamefont {Li}}, \bibinfo {author} {\bibfnamefont {X.}~\bibnamefont {Ma}}, \bibinfo {author} {\bibfnamefont {T.}~\bibnamefont {Kuang}}, \bibinfo {author} {\bibfnamefont {H.~P.}\ \bibnamefont {Zhang}},\ and\ \bibinfo {author} {\bibfnamefont {W.}~\bibnamefont {Wang}},\ }\href {https://doi.org/10.1039/D5SM00073D} {\bibfield  {journal} {\bibinfo  {journal} {Soft Matter}\ }\textbf {\bibinfo {volume} {21}},\ \bibinfo {pages} {3515} (\bibinfo {year} {2025})}\BibitemShut {NoStop}%
\bibitem [{\citenamefont {Campbell}\ and\ \citenamefont {Ebbens}(2013)}]{campbell_gravitaxis_2013}%
  \BibitemOpen
  \bibfield  {author} {\bibinfo {author} {\bibfnamefont {A.~I.}\ \bibnamefont {Campbell}}\ and\ \bibinfo {author} {\bibfnamefont {S.~J.}\ \bibnamefont {Ebbens}},\ }\href {https://doi.org/10.1021/la403450j} {\bibfield  {journal} {\bibinfo  {journal} {Langmuir}\ }\textbf {\bibinfo {volume} {29}},\ \bibinfo {pages} {14066} (\bibinfo {year} {2013})}\BibitemShut {NoStop}%
\bibitem [{\citenamefont {Lyu}\ \emph {et~al.}(2021)\citenamefont {Lyu}, \citenamefont {Liu}, \citenamefont {Zhou}, \citenamefont {Duan}, \citenamefont {Xu}, \citenamefont {Dai}, \citenamefont {Chen}, \citenamefont {Peng}, \citenamefont {Cui}, \citenamefont {Tang}, \citenamefont {Ma},\ and\ \citenamefont {Wang}}]{lyu_active_2021}%
  \BibitemOpen
  \bibfield  {author} {\bibinfo {author} {\bibfnamefont {X.}~\bibnamefont {Lyu}}, \bibinfo {author} {\bibfnamefont {X.}~\bibnamefont {Liu}}, \bibinfo {author} {\bibfnamefont {C.}~\bibnamefont {Zhou}}, \bibinfo {author} {\bibfnamefont {S.}~\bibnamefont {Duan}}, \bibinfo {author} {\bibfnamefont {P.}~\bibnamefont {Xu}}, \bibinfo {author} {\bibfnamefont {J.}~\bibnamefont {Dai}}, \bibinfo {author} {\bibfnamefont {X.}~\bibnamefont {Chen}}, \bibinfo {author} {\bibfnamefont {Y.}~\bibnamefont {Peng}}, \bibinfo {author} {\bibfnamefont {D.}~\bibnamefont {Cui}}, \bibinfo {author} {\bibfnamefont {J.}~\bibnamefont {Tang}}, \bibinfo {author} {\bibfnamefont {X.}~\bibnamefont {Ma}},\ and\ \bibinfo {author} {\bibfnamefont {W.}~\bibnamefont {Wang}},\ }\href {https://doi.org/10.1021/jacs.1c04501} {\bibfield  {journal} {\bibinfo  {journal} {J. Am. Chem. Soc.}\ }\textbf {\bibinfo {volume} {143}},\ \bibinfo {pages} {12154} (\bibinfo {year} {2021})}\BibitemShut {NoStop}%
\bibitem [{\citenamefont {Park}\ \emph {et~al.}(2019)\citenamefont {Park}, \citenamefont {Na},\ and\ \citenamefont {Bae}}]{park_uniform_2019}%
  \BibitemOpen
  \bibfield  {author} {\bibinfo {author} {\bibfnamefont {B.}~\bibnamefont {Park}}, \bibinfo {author} {\bibfnamefont {S.~Y.}\ \bibnamefont {Na}},\ and\ \bibinfo {author} {\bibfnamefont {I.-G.}\ \bibnamefont {Bae}},\ }\href {https://doi.org/10.1038/s41598-019-47990-z} {\bibfield  {journal} {\bibinfo  {journal} {Sci Rep}\ }\textbf {\bibinfo {volume} {9}},\ \bibinfo {pages} {11453} (\bibinfo {year} {2019})}\BibitemShut {NoStop}%
\bibitem [{\citenamefont {Wei}\ \emph {et~al.}(2018)\citenamefont {Wei}, \citenamefont {Zhou}, \citenamefont {Tang},\ and\ \citenamefont {Wang}}]{Surface_charges_wang_2018}%
  \BibitemOpen
  \bibfield  {author} {\bibinfo {author} {\bibfnamefont {M.}~\bibnamefont {Wei}}, \bibinfo {author} {\bibfnamefont {C.}~\bibnamefont {Zhou}}, \bibinfo {author} {\bibfnamefont {J.}~\bibnamefont {Tang}},\ and\ \bibinfo {author} {\bibfnamefont {W.}~\bibnamefont {Wang}},\ }\href {https://doi.org/10.1021/acsami.7b18399} {\bibfield  {journal} {\bibinfo  {journal} {ACS Applied Materials \& Interfaces}\ }\textbf {\bibinfo {volume} {10}},\ \bibinfo {pages} {2249} (\bibinfo {year} {2018})},\ \bibinfo {note} {pMID: 29300455},\ \Eprint {https://arxiv.org/abs/https://doi.org/10.1021/acsami.7b18399} {https://doi.org/10.1021/acsami.7b18399} \BibitemShut {NoStop}%
\end{thebibliography}
\end{document}